\documentclass[prl,showpacs,twocolumn,groupedaddress,floatfix,a4paper,10point]{revtex4-1}

\usepackage{epsfig}
\usepackage[T1]{fontenc}
\usepackage{graphicx}
\usepackage{amssymb}
\usepackage{amsmath}
\usepackage{relsize}
\usepackage{color}

\usepackage{subfig}

\begin{document}

\title{Reacting particles in open chaotic flows}

\author{Alessandro P. S. de Moura}
\affiliation{Institute of Complex Systems and Mathematical Biology,
  University of Aberdeen, Aberdeen, AB24 3UE, UK}

\begin{abstract}
  We study the collision probability $p$ of particles advected by open
  flows displaying chaotic advection. We show that $p$ scales with the
  particle size $\delta$ as a power law whose coefficient is
  determined by the fractal dimensions of the invariant sets defined
  by the advection dynamics. We also argue that this same scaling also
  holds for the reaction rate of active particles in the low-density
  regime. These analytical results are compared to numerical
  simulations, and we find very good agreement with the theoretical
  predictions.
\end{abstract}

\pacs{ 
  47.52.+j,
  47.53.+n,
  05.45.Ac,
  47.51.+a
}

\maketitle

\begin{figure}[t]
  \centering
  \includegraphics[width=0.4\textwidth]{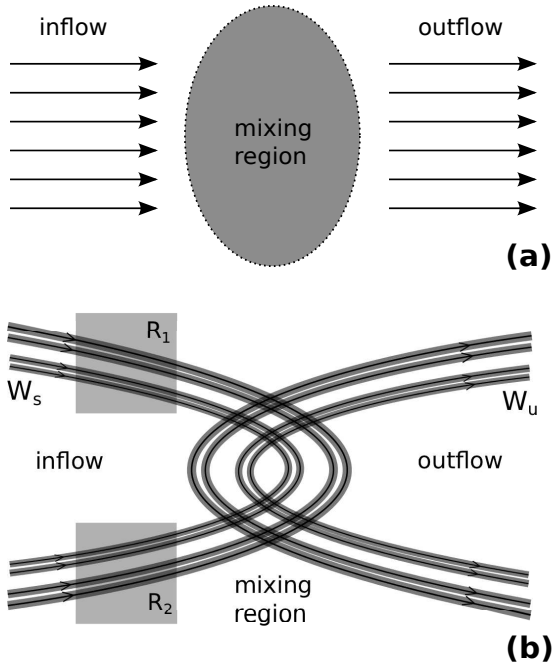}
  \caption{(a) Illustration of an open flow. (b) Schematic depiction
    of the stable ($W_s$) and unstable ($W_u$) manifolds in an open
    flow. The grey bands are $\delta$-neighbourhoods of $W_s$ and
    $W_u$ (see text). $R_1$ and $R_2$ indicate regions where
    initial conditions are chosen.}
  \label{fig:flow}  
\end{figure}

Many fluid flows of interest to science and to engineering are open
flows, which are characterised by the presence of inflow and
outflow regions \cite{batchelor} (see Fig. \ref{fig:flow}a).  The
dynamics of advected particles in open flows is transient, with
typical particles leaving to the outflow region in finite time.
Advection in many open flows is chaotic
\cite{Jung-et-al-93,Tel-et-al-05}; examples of chaotic open flows are
found in areas such as microfluidics \cite{Strook2002}, climatology
\cite{Koh2002}, physiology \cite{Schelin2009}, population biology
\cite{Karolyi2000} and industry
\cite{Gouillart2009a,Gouillart2011measures}.  The dynamics of advected
particles is governed by the \emph{chaotic saddle} \cite{cscatbook},
which is a set of unstable orbits contained in a bounded region of
space known as the \emph{mixing region} (see Fig. \ref{fig:flow}a).
Fluid elements are repeatedly stretched and folded in the mixing
region, and this causes any portion of the fluid to be deformed by the
flow into a complex filamentary shape, which shadows the
\emph{unstable manifold} of the chaotic saddle, defined as the fractal
set of orbits which converge asymptotically to the chaotic saddle for
$t\rightarrow-\infty$ \cite{cscatbook}.

The fractal structures generated by chaotic advection have dramatic
consequences for the dynamics of active processes taking place in open
flows, such as chemical reactions and biological processes
\cite{Tel-et-al-05}.  The stretching and folding of fluid elements by
the flow tends to increase the area (or perimeter, in 2D flows) of
contact between two reacting species, which results in an enhancement
of the reaction due purely to the advective dynamics of the
flow---this has been called \emph{dynamic catalysis}
\cite{Tel-et-al-05}.  It has been shown
\cite{Karolyi1999,Tel-et-al-05} that this effect appears as a singular
production term in an effective reaction rate equation, which has a
power-law dependence on the amount of reactant; the coefficient of the
power law is determined by the information fractal dimension of the
unstable manifold of the chaotic saddle.  Later works have established
that this enhancement of activity by chaos is a very general and
robust phenomenon \cite{Tel-et-al-05}, and is a feature of many kinds
of activity and flows, including non-periodic \cite{Karolyi2004},
non-hyperbolic \cite{Motter2003,Moura2004} and three-dimensional
\cite{Moura2004b} flows.

All the works so far on the effects of chaos on chemical or biological
activity in open flows have adopted continuum descriptions of the
reacting material --- either describing the reactants using continuous
concentration fields, or modelling the propagation of the reaction
through reaction fronts.  This approach ignores the fact that any
activity taking place in the fluid ultimately arises from collisions
of particles.  If the number of reacting particles is very large, so
that the mean free path between collisions is much smaller than the
typical length scale of the system, then this continuous description
is expected to be valid.  However, in the \emph{low-density limit},
where typically particles traverse large distances before colliding
and reacting, the concepts of reactant concentration and reaction
front have no meaning.  In this case, one must adopt a kinetic theory
approach, where activity is described in terms of the probability of
collision between particles.  Although this idea has been used in
closed flows --- in the context of chemical catalysis
\cite{Metcalfe1994}, particle coalescence
\cite{Nishikawa2001,Springer2010} and crystallisation
\cite{Cartwright2004}, for example --- so far this idea has not been
pursued in the case of open flows, despite the importance of the
low-density regime for applications, which is appropriate for
describing systems as diverse as platelet activation in blood flows
\cite{Schelin2009}, plankton population dynamics \cite{Karolyi2004},
and raindrop formation \cite{Wilkinson2006,Vilela2007}.

In this work we investigate the effect of chaotic advection on the
activity of particles in the low-density limit.  The rate of reaction
in this limit is determined by the collision probability $p(\delta)$
of two particles coming within a distance $\delta$ of each other
before they escape to the outflow; we assume a reaction event takes
place when that happens.  $\delta$ can be thought of as the size of
the particles, or alternatively as determined by the reaction
cross-section $\sigma$ (with $\delta\sim\sqrt\sigma$).  We show that
$p$ scales with $\delta$ as a power law,
$p(\delta)\sim\delta^{\beta}$, and we derive an analytical expression
for the coefficient $\beta$ in terms of the fractal dimensions of the
stable and unstable manifolds of the chaotic saddle.  The classical
kinetic theory approach, based on the Smoluchowski equation
\cite{Smoluchowski1916}, which assumes there is a homogeneous mixing
of particles in the mixing region, would predict $\beta=N$, where $N$
is the spatial dimension, but we show that because most collisions
happen in the neighbourhood of a fractal set, Smoluchowski's result
does not hold, and $\beta\neq N$ for open chaotic flows.  In the
low-density limit, the total number $Q(\delta)$ of reaction events
between reacting particle species which takes place in a given flow
also scales as $\delta^{\beta}$.  We argue that $Q$ is a good measure
of the efficiency of mixing in open flows, and thus $\beta$
characterises the mixing of an open flow. These analytical predictions
are compared to numerical simulations in 2D flows, and we find the
results agree very well with our theory.

We start by deriving the scaling of the collision probability
$p(\delta)$ of two particles with randomly chosen initial conditions,
by first assuming that after the particles enter the mixing region
(Fig. \ref{fig:flow}a), the flow acts as a perfect, uniform mixer,
until the particles escape to the outflow.  This perfect mixing
assumption means that the positions of the two particles are
effectively randomised in a time interval $\tau$ which is
characteristic of the flow.  The probability that two particles
randomly placed in a bounded $N$-dimensional region are within a
distance $\delta$ of each other scales as $\delta^N$.  Assuming that
$\delta$ is small enough so that the probability $p(\delta)$ that the
two particles collide before escaping satisfies $p(\delta)\ll 1$, we
conclude that $p(\delta)$ is proportional to $\delta^N$ and to the
average time $T$ the particle stays in the mixing region before
escaping.  $p(\delta)$ therefore scales with $\delta$ as
\begin{equation}
\label{smoluch}
p(\delta) \sim \delta^N,
\end{equation}
The perfect mixing assumption is the assumption used in Smoluchowski's
classical coagulation theory \cite{Smoluchowski1916}, and we will
refer to this $\delta^N$ scaling as the Smoluchowski prediction.

The Smoluchowski scaling derived above does not take into account the
fractal nature of the particle distributions generated by chaotic
advection in open flows.  We will now derive the correct scaling of
$p(\delta)$.  Let two particles be chosen randomly within two bounded
and disjoint regions $R_1$ and $R_2$ (see Fig. \ref{fig:flow}b).  As a
result of the chaotic dynamics of the flow in the mixing region, those
trajectories which take longer to escape converge towards the unstable
manifold $W_u$.  Consequently, the particles will only have a chance
to meet if both come within a distance $\delta$ (the reaction
distance) of the unstable manifold.  This will take place if their
initial conditions are within a distance of order $\delta$ of the
stable manifold $W_s$, which is the set of initial conditions
converging to the chaotic saddle for $t\rightarrow\infty$ (see
Fig. \ref{fig:flow}b) \cite{cscatbook}.  The probability
$P_s^{(1)}(\delta)$ that one particle with initial conditions in $R_1$
lies within a distance of order $\delta$ of the stable manifold is
proportional to the volume $V_1$ (or area in two dimensions) of the
intersection of $R_1$ and a $\delta$-covering of the component of
$W_s$ where the conditionally invariant measure is concentrated
\cite{Grassberger1983,cscatbook} (see Fig. \ref{fig:flow}b).  If one
imagines $R_1$ to be covered by a grid of size $\delta$, the number of
cells in the grid intersecting the component of $W_s$ where the
conditionally invariant measure is concentrated scales as
$N_1(\delta)\sim\delta^{-D_1^s}$ \cite{cscatbook}, where $D_1^s$ is
the information dimension of the stable manifold.  Since the volume of
each cell in the grid is $\delta^N$, the volume of the portion of the
grid which intersects $W_s$ scales as
$V_1\sim N(\delta)\delta^{N}\sim\delta^{N-D_1^s}$.  We therefore have
$P_s^{(1)}(\delta)\sim\delta^{N-D_1^s}$, and the same scaling holds
for the probability $P_s^{(2)}$ that the other particle is close to
$W_s$.  Therefore, the probability $P_s=P_s^{(1)}P_s^{(2)}$ that both
particles reach a $\delta$-neighbourhood of the unstable manifold
scales as
\begin{equation}
P_s(\delta) \sim \delta^{2N-2D_1^s}.
\label{Ps}
\end{equation}
The two particles collide if they reach the unstable manifold within a
distance $\delta$ of each other.  Considering again a grid of size
$\delta$ covering the unstable manifold $W_u$, the probability
$P_u(\delta)$ that two particles collide can be approximated by the
probability that they end up in the same cell in the grid.  This is
given by $P_u(\delta)=\sum_i\mu_i^2$, where $\mu_i$ is the
(conditionally invariant) measure of cell $i$, and the sum is over all
cells intersecting the unstable manifold.  From the definition of the
correlation dimension $D_2^u$ \cite{Grassberger1983,cscatbook}, we
find that in the limit $\delta\rightarrow 0$ we get
\begin{equation}
P_u(\delta) \sim \delta^{D_2^u}.
\label{Pu}
\end{equation}
The overall probability $p(\delta)=P_s(\delta)P_u(\delta)$ that the
two particles will collide therefore scales as
\begin{equation}
p(\delta) \sim \delta^{\beta},
\hspace{0.5cm}
\text{with}\ \  \beta=2N-2D_1^s+D_2^u.
\label{prob}
\end{equation}
This is our main result.  This scaling is clearly different from the
Smoluchowski prediction: for chaotic flows, in general we expect
$\beta\neq N$.  However, $\beta$ does approach the Smoluchowski value
of $N$ in the limit where the escape rate vanishes, when $D_1^2, D_2^u
\rightarrow N$, and from Eq. (\ref{prob}) this implies that
$\beta\rightarrow N$.  This behaviour is expected, since in this limit
the dynamics within the mixing region can be considered as that of a
closed container being stirred, with a small leak, when particles
spend a long lime in the mixing region and the perfect mixing
assumption is approximately valid.

The scaling predicted by Eq. (\ref{prob}) is valid for any choice of
regions $R_1$ and $R_2$, as long as both regions intersect the stable
manifold $W_s$. If there is no intersection, the two particles never
meet because they do not get to the mixing region, and collisions
cannot take place.

\begin{figure}[t]
  \centering
  \includegraphics[width=0.4\textwidth,angle=0]{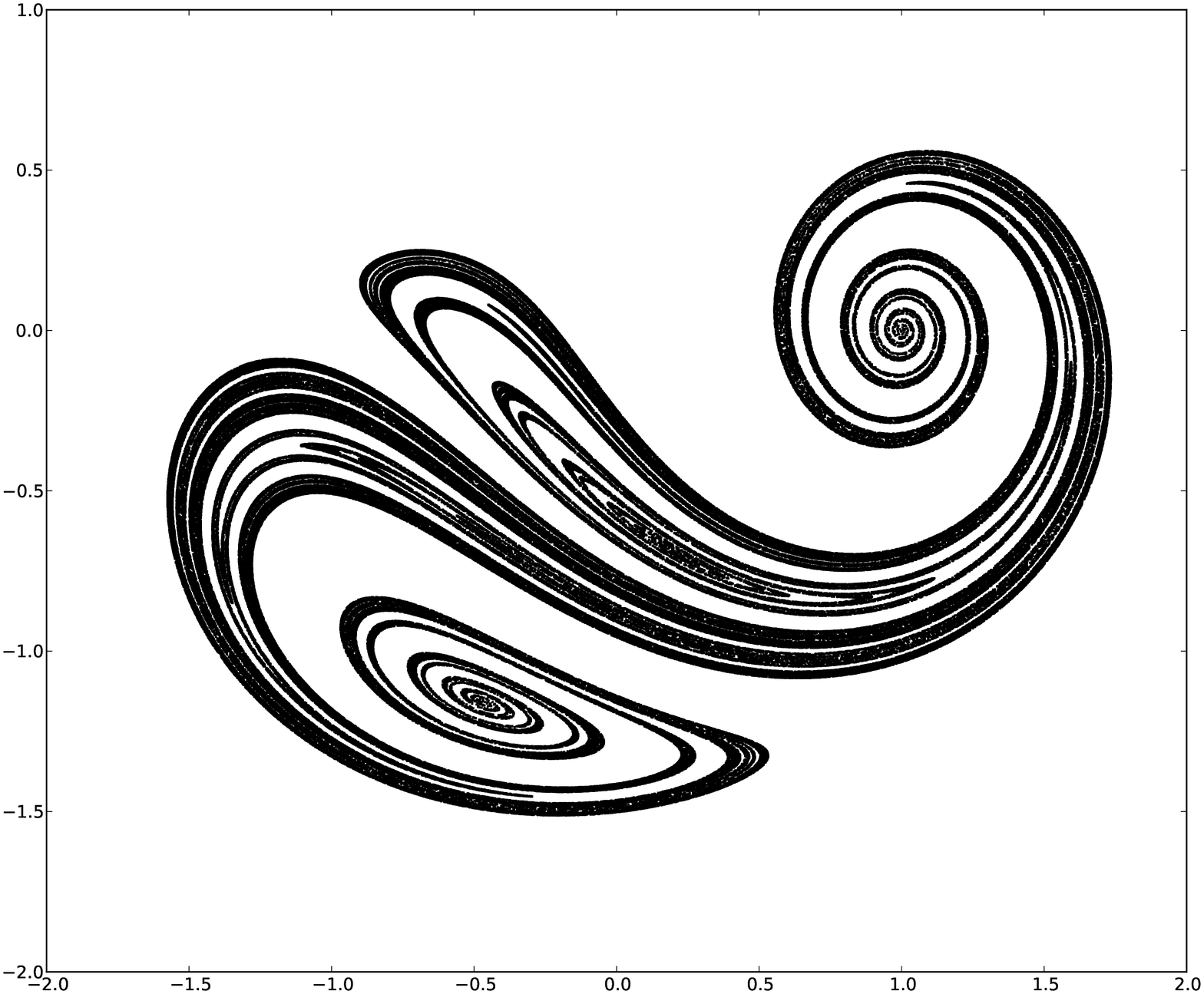}
  \includegraphics[width=0.4\textwidth,angle=0]{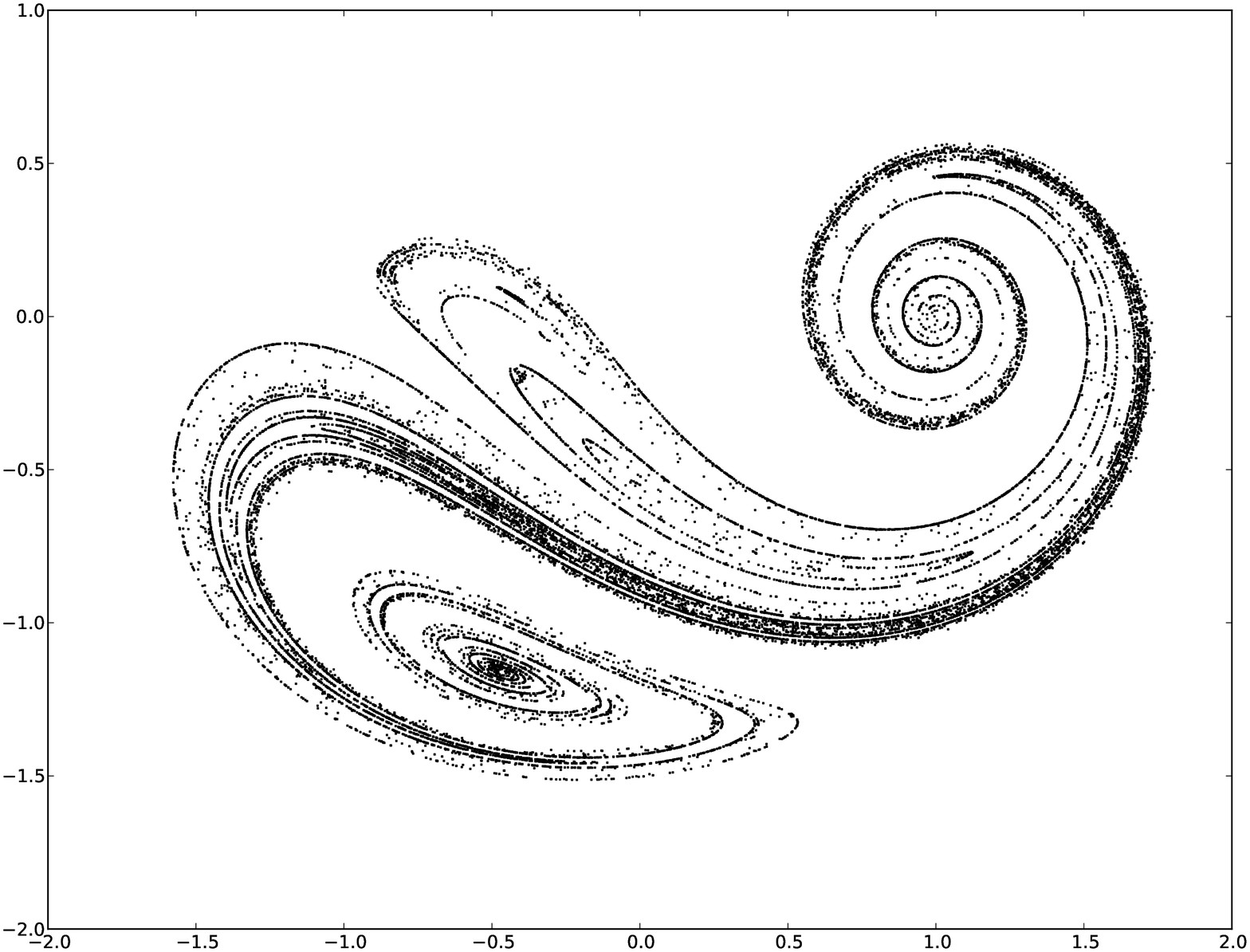}
  \caption{(a) Numerical approximation of the unstable manifold for
    the blinking vortex-sink flow, with $\eta=0.5$ and $\xi=10$; for
    these parameters, the dynamics is chaotic and hyperbolic. (b)
    Location of collisions for $\delta=10^{-3}$, for the flow with the
    same parameters as above.  It is clear that collisions take place
    in the vicinity of the unstable manifold.  The initial conditions
    were chosen in the regions $R_1$ ($-0.5<x<0.5$, $0<y<0.15$) and
    $R_2$ ($-0.5<x<0.5$, $0.2<y<0.35$). The time step used in the
    simulations was $\Delta t=0.02$.}
  \label{fig:blink}
\end{figure}

To test the prediction of Eq. (\ref{prob}), we will use the
\emph{blinking vortex-sink flow}, \cite{PHR,Aref1989bs} which is a
generalisation of Aref's blinking vortex flow \cite{Aref1984}.  It is
a 2D incompressible flow on an infinite plane, with two sinks at
positions $(+a,0)$ and $(-a,0)$, which open and close periodically in
alternation: in the first half of each cycle one sink is open and the
other one is closed, and in the second half the situation is reversed.
Each vortex-sink is modelled as a point source of vorticity
superimposed to a localised sink, and flow which falls on either of
the sinks disappears from the system and does not come back. This
system is clearly an open flow, where the inflow region corresponds to
the whole space beyond the sinks.

%

\begin{figure}[t]
  \centering
  \includegraphics[width=0.4\textwidth]{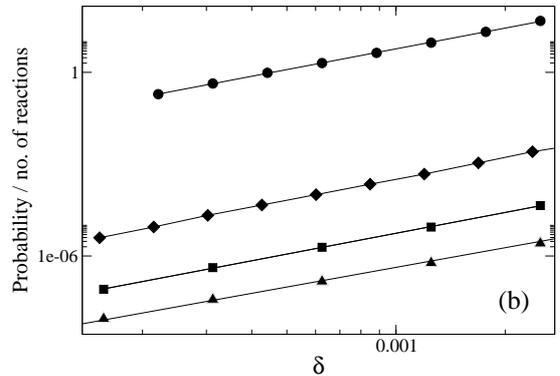}
  \caption{ Collision probability (squares) and the number of reaction
    events (circles) as a function of the reaction distance $\delta$
    for the blinking vortex-sink flow, with the same parameters as in
    Fig. \ref{fig:blink}. The initial conditions were chosen in the
    regions $R_1$ and $R_2$ defined in Fig. \ref{fig:blink}.  The
    number of pairs used varied with $\delta$, and it was such that it
    guaranteed that at least $10^3$ collisions took place for each
    $\delta$; typically this meant the number of pairs was of order
    $10^{10}$--$10^{11}$.  The straight lines are the fitting of the
    data to a power law of the form $\delta^{\beta^*}$; in both cases
    fitting yields $\beta^*=2.23$, compared to the prediction
    $\beta=2.25$ of Eq. (\ref{prob}).  The collision probability for
    the area-preserving baker map with escape is also shown
    (diamonds), for contraction factors $0.3$ and $0.45$ \cite{Ott}.
    Fitting yields $\beta^*=2.30$.  For this system $D_1^s=1.701$ and
    $D_2^u=1.698$, and from Eq. (\ref{prob}) the prediction is
    $\beta=2.298$.  The triangles show the collision probability for
    the conservative H\'enon map for the nonlinearity parameter
    $\lambda=7$ \cite{Moura2004}, for which $D_1^s=1.58$ and
    $D_2^u=1.32$, leading to a prediction of $\beta=2.16$; fitting
    yields $\beta^*=2.14$.  The baker map data (diamonds) in this plot
    has been shifted in the graph above to improve visibility. }
  \label{fig:scaling}
\end{figure}

Assuming that the particles advected by the flow behave as passive
tracers, their equations of motion are \cite{PHR}
\begin{equation}
\label{flow}
\dot{r} =-C/r, \quad \dot{\varphi}=K/r^2,
\end{equation}
where $r$ and $\varphi$ are polar coordinates whose origin alternates
between $(a,0)$, during the first half of each period, and $(-a,0)$,
during the second half of each period; $C$ and $K$ are parameters
representing the strengths of the sink and the vortex, respectively.
The advection dynamics is determined by the two dimensionless
parameters $\eta=CT/ a^2$ and $\xi=K/C$, where $T$ is the flow period.
We choose units so that $a=1$ and $T=1$.  The dynamics is chaotic for
a wide range of $\eta$ and $\xi$ \cite{PHR}.  Fig. \ref{fig:blink}a
shows the unstable manifold for one particular choice of parameters,
$\eta=0.5$ and $\xi=10$.

We computed the collision probability $p(\delta)$ by following the
trajectories of many pairs of particles until one particle in each
pair escapes, and at every time step $\Delta t$ we check if the
distance between the two particles is less than $\delta$; if so, the
particles are considered to have collided.  The initial condition of
one of the particles in each pair is chosen randomly within region
$R_1$, defined in the caption of Fig. \ref{fig:blink}; and the other
particle in the pair is started in region $R_2$ (see
Fig. \ref{fig:blink}).  The fraction of the pairs which collide before
escaping gives us a numerical approximation of $p(\delta)$.  Figure
\ref{fig:scaling} (the squares) shows the result of applying this
procedure for a range of values of $\delta$.  The resulting scaling is
clearly a power law $p(\delta)\sim\delta^{\beta^*}$, and fitting
yields $\beta^*\approx 2.25$.  To compare this with the prediction of
Eq. (\ref{prob}), we use the value $D_1^s\approx 1.74$ from
\cite{PHR}; to compute $D_2^u$ we first find an approximation
$W_u^{\rm spr}$ of the unstable manifold using the sprinkler method
\cite{cscatbook}, and then we compute the scaling with $\epsilon$ of
the total number of pairs of points on $W_u^{\rm spr}$ separated by a
distance less than $\epsilon$; the coefficient of the resulting power
law is $D_2^u$ \cite{Grassberger1983}.  We found $D_2^u\approx 1.71$.
This is a two-dimensional system, and so $N=2$ in
Eq. (\ref{prob}). Equation (\ref{prob}) then predicts $\beta=2.23$,
which agrees with the value $\beta^*=2.25$ from the simulation, to
within numerical errors.

Fig. \ref{fig:blink}a is a snapshot of $W_u$ at the start of a period;
the shape of $W_u$ changes periodically with time, with the same
period as the flow.  Figure \ref{fig:blink}b shows the positions of
the collisions that take place at the beginning of each period: each
dot in Fig. \ref{fig:blink}b is one such collision.  Comparing
Figs. \ref{fig:blink}a and \ref{fig:blink}b, it is clear that the
collisions take place in a small neighbourhood of the unstable
manifold $W_u$, confirming that the assumption made in the derivation
of Eq. (\ref{prob}) is correct.

We have analysed other choices of the parameters $\eta$ and $\xi$ for
which the flow is chaotic, and we always find that the prediction from
Eq. (\ref{prob}) and the results from the simulations differ by
less than 1\% in all cases.  We found that the choice of the time step
$\Delta t$ does not affect the scaling of $p(\delta)$ with $\delta$
--- although the value of $p$ for any given $\delta$ of course
decreases with $\Delta t$, since with larger time steps it is more
likely collisions will be missed by the simulation.

We have also studied the dynamics of collisions in other
two-dimensional dynamical systems, including a variant of the
conservative H\'enon map which displays transient chaos
\cite{Moura2004}, and a version of the baker's map with escape
\cite{Ott}.  In all cases, and for all parameter choices, we found
excellent agreement between the numerical value of $\beta$ and the
prediction of Eq. (\ref{prob}) (see Fig. \ref{fig:scaling}).

We now examine a system of reacting particles advected by a chaotic
open flow.  Taking as an illustrative example the case of an
acid-base-type reaction $A+B\rightarrow C$, consider particles of two
species, which we will label simply species $A$ and $B$, such that
when an $A$ particle comes within a distance $\delta$ of a particle of
type $B$, a reaction occurs, resulting in a particle of type $C$,
which is the product of the reaction.  If we throw $N_1$ type-$A$
particles in a region $R_1$ of the flow, and $N_2$ type-$B$ particles in
region $R_2$, then we expect that the total number $N_p$ of type-$C$
particles produced in the system is proportional to $N_1N_2p(\delta)$
--- as long as $N_1$ and $N_2$ are not too large, and the low-density
assumption holds.  We therefore predict that the amount $N_p$ of
product generated by the reaction scales with the reaction distance
$\delta$ as $\delta^\beta$, with $\beta$ given by Eq. (\ref{prob}).

We put this prediction to the test in the blinking vortex-sink system,
using an efficient algorithm for finding the neighbours within a
distance $\delta$ \cite{Bentley1975}.  The resulting scaling of $N_p$
with $\delta$ is shown in Fig. \ref{fig:scaling}.  We find a power-law
scaling as expected, and fitting yields the coefficient
$\beta^*=2.25$, which is again within 1\% of the predicted value of
$\beta=2.23$.  We found the same agreement for other choices of
parameters for the blinking vortex-sink system, and also for the other
dynamical systems we investigated.

We note that, even though we used the reaction $A+B\rightarrow C$ as
an example above, all kinds of reactions and active processes
involving particles will be affected by the scaling (\ref{prob}),
since all reactions require particles to come close together, and the
probability of that happening is governed by the coefficient $\beta$
of Eq. (\ref{prob}).  We propose that the collision probability $p$,
and the way is scales with the collision distance $\delta$, is the
natural quantity to characterise the efficiency of mixing of an open
flow (an alternative definition is given in
\cite{Gouillart2011measures}).  Equation (\ref{prob}) shows that the
mixing efficiency is determined by the fractal properties of the
invariant sets associated with the chaotic saddle --- namely its
stable and unstable manifolds.  The discussion above shows that the
mixing efficiency in turn determines the efficiency of chemical
reactions (or, in general, active processes) in open flows, which
means that the ``chemical efficiency'' of a flow is also measured by
the coefficient $\beta$ defined in Eq. (\ref{prob}).

\bibliography{collision_paper}

\end{document}